**Classification:** Major category: Biological Sciences; minor category: Evolution

**Title: The hypothesis that coelacanth is the closest living relative of tetrapods was rejected based on three genome-scale approaches**

**Authors:** Yunfeng Shan*†‡, Xiu-Qing Li† and Robin Gras‡

**Author affiliation:** *Department of Natural History, Royal Ontario Museum, 100 Queens' Park, Toronto, Ontario M5S 2C6, Canada; †Molecular Genetics Laboratory, Potato Research Centre, Agriculture and Agri-Food Canada, 850 Lincoln Rd, P.O. Box 20280, Fredericton, New Brunswick, E3B 4Z7, Canada; ‡School of Computer Science, University of Windsor, 401 Sunset Avenue, Windsor, ON  N9B 3P4, Canada

**Corresponding author:**  Yunfeng Shan

**Manuscript information:**

The number of text pages: 18

The number of figures: 2

The number of tables: 3

______________________

Abbreviations: ML, maximum likelihood; NJ, neighbor joining; MP, maximum parsimony; M, Mammal; B, Bird; A, Amphibian; C, Coelacanth; L, Lungfish: R, Ray-finned Fish; S, Shark.

**Since its discovery of the "living fossil" in 1938, the coelacanth (*Latimeria chalumnae*) has generally been considered to be the closest living relative of the land vertebrates, and this is still the prevailing opinion in most general biology textbooks. However, the origin of tetrapods has been the subject of intense debate for decades. The three principal hypothesis (lungfish-tetrapod, coelacanth-tetrapod, or lungfish-coelacanth sister group) have been proposed. We used the maximum gene-support tree approach to analyze 43 nuclear genes encoding amino acid residues, and compared the results of concatenation and majority-rule tree approaches. The results inferred with three common phylogenetic methods and three genome-scale approaches consistently rejected the hypothesis that the coelacanth is the closest living relative of tetrapods.**

T he origin of land vertebrates (tetrapods) has been subject to debate for many decades, yet the relationship of the extant basal taxa remains contentious. The origin of the tetrapod is of importance and always commands considerable popular interests in public and academic fields. However, since the discovery of the "living fossil" in 1938, *Latimeria chalumnae* (1, 2), the last known surviving species of a lineage of lobe-finned fish, was generally considered to be the closest living relative of the land vertebrates, the missing link between aquatic and terrestrial vertebrates. This hypothesis is still the prevailing opinion in most general biology textbooks (3). Three hypotheses have been proposed for the phylogenetic relationship: e.g., lungfish-tetrapod (hypothesis 1, Fig 1a), coelacanth-tetrapod (hypothesis 2, Fig 1b), or lungfish-coelacanth sister grouping (hypothesis 3, Fig 1c). The lungfish-coelacanth-tetrapod trichotomy (Fig 1d) is not generally considered as a hypothesis.

**Fig. 1.**

Regarding which of the three extant basal taxa was the closest living relative to tetrapods, the coelacanth hypothesis (Fig 1b) was supported by some comparative morphologists and paleontologists (4-7), although the lungfish was historically thought to have that claim, which was also supported by more recent researchers (8-12). The hypothesis that coelacanths and lungfishes form a monophyletic group that is equally closely related to the tetrapods (Tree III) was also proposed (13-15).

Molecular data have gradually been collected to infer phylogenetic relationship for the last two decades. With mitochondrial DNA, lungfishes were linked as the closest relatives of tetrapods by single genes (16-25); the coelacanth as the closest living sister group of tetrapods was preferred by 12S and 16S rRNA genes although this hypothesis received the least support in all phylogenetic analysis of molecular data (3), and the coelacanth and lungfish sister group relationship was suggested by the single gene (19) and the whole genome (23, 25), while an unresolved coelacanth-lungfish-tetrapod trichotomy resulted by 12S rRNA gene (18). Nuclear gene-based phylogenetic reconstructions also generated the first three options that the closest relative of tetrapods would be either lungfishes (26-28), coelacanths (29-32), or equally to both coelacanths and lungfishes that form a sister group (21).

Recently, this question was re-investigated using 44 genes with a concatenation genome-scale approach (33). However, the result was an unresolved trichotomy.

Although many morphology and molecule based phylogentic studies have attempted to resolve this question, the results mentioned above so far do not discover unequivocal evidence as to whether the coelacanth or the lungfish is the closest living relatives of tetrapods or both lineages are equally closely related to tetrapods. Therefore, the origin of tetrapods continued to be debated and still is one of the longest standing major questions in vertebrate evolution.

Significant progress in genome sequencing technology and recent completed genome projects results in huge amounts of sequence data from a lot of organisms, which results in expectations that in the near future, recovering the tree of life (TOL) will simply be a matter of enough sequence data collection with concatenated multiple gene approach. However, recent concatenated multiple gene analyses of some key clades in life's history have not resolved phylogenetic trees due to homoplasy (34). Alternative methodology is still necessary.

It is clear that not all phylogenetic methods and genome-scale approaches are equally powerful and reliable (28); however, it is generally accepted that when several phylogenetic methods and approaches converge on the same topology, that can be taken as added evidence in support of a particular hypothesis (28).

In order to resolve the origin question of tetrapods, we used maximum gene-support, an alternative genome-scale approach (35), and compared it with two other genome-scale approaches to analyze all 43 nuclear genes encoding amino acid residues that are currently available in GenBank with all three common phylogentic methods. We sampled 7 taxa: Mammal (M), Bird (B), Amphibian (A), Coelacanth (C), Lungfish (L), Ray-finned Fish (R), and Shark (S).

**Results**

**Seven taxon (MBACLRS) set.** Gene support is the number of genes that reconstructed a unique topology. As shown in Table 1, gene supports were equal, e.g., 2, for all four tree types with the maximum parsimony (MP) method. Four tree types of 7 taxa were shown in Fig. 2a-d. No unique maximum gene support tree was identified. Therefore, this phylogeny was irresolvable using maximum gene-support tree approach for these 7 taxa with MP. The irresolvable results were also observed with the maximum likelihood (ML) and neighbor-joining (NJ) methods (Table 1)

**Fig. 2.**

Table 1

Phylogenetic analysis with these three common phylogenetic methods and three approaches did not converge as shown in Table 1. Results evidently varied with the methods and the approaches. The lungfishes as the closest living relatives was inferred with ML with 100% bootstrap support, but the lungfishes and coelacanth sister group was recovered with NJ with 87% support with concatenated multiple gene approach. The maximum gene support tree approach clearly showed that 43 genes were not able to resolve the phylogenetic relationship for these 7 taxa regardless of the phylogentic methods used.

**Six taxon sets**. Table 2 showed tree II was supported by many fewer genes than tree I or tree III for all three methods. Significant differences in gene supports of tree II and tree I or tree III inferred with MP were observed for MBACLR and MACLRS (Table 2) at $P < 0.10$ level by means of chi-square test. There were no significant differences in the gene supports of tree I and tree III. Tree IV was supported by one gene for the taxon set of MBCLRS only.

Table 2

**Five taxon sets.** Chi-square test showed that significant lower gene support of tree II than tree III was observed for MBCLR at $P < 0.05$ significant level with MP. Significant lower gene supports of tree II than tree I for BACLS at $P < 0.05$, MACLS at $P < 0.10$ and MBCLS at $P < 0.10$ with ML were detected (Table 2). There were no significant differences in gene supports between Tree I and tree III of all nine 5-taxa sets (Table 2).

**Four taxon sets.** Tree II was inferred with only taxon set of ACLS with MP and maximum gene support tree approach (Table 2). Significant lower gene supports of tree II than tree III or I were observed for ACLR and BCLS at $P < 0.05$ significant level with NJ (Table 2), and for BCLS at $P < 0.05$ level with ML (Table 2) based on Chi-square test. No significant differences between tree I and tree III were observed in gene supports.

The taxon jackknife analysis (Table 3) showed that jackknife probability was 10.0% for tree II, 27.5% tree I and 62.5% for tree III with maximum gene-support tree approach and MP. Zero probability for tree II, 40% jackknife probability for tree I and 50% for tree III were observed with the concatenated genome-scale approach and MP. Jackknife probability was 10% for tree II, 30% for tree I, and 60% for tree III with maximum gene-support branches approach. The jackknife probability of tree IV was zero for three approaches (Table 2) and MP. Similar results were observed for ML and NJ (Table 3).

**Discussion**

When 43 genes are used to reconstruct the phylogeny of 7 taxa, the maximum gene-support tree approach gives no resolution. The maximum gene-support branch approach infers tree III by MP and ML, and tree II by NJ. The concatenation approach recovers tree III with NJ, and tree I with MP and ML. These results vary with phylogenetic methods and are very inconsistent. The results of the maximum gene-support tree approach clearly show that 43 genes do not reach the threshold of the minimum number of genes required for resolution of the phylogeny of these 7 taxa. Because the number of alternative trees increases exponentially with the number of taxa (6→945, 7→10,395, n→ (2n-3)!!) (36), minimum of required genes increases with increased taxon number in a taxon set. In order to meet the minimum requirement of genes, one way is to increase the genes, while another way is to decrease the taxa. Here, we choose the latter although the debate of taxon sampling. We use a jackknife approach to sub-sample 6, 5 and 4 taxa from 7 taxa each time to reduce taxon number and subsequently to decrease alternative trees.

How many genes are required? For recent years, concatenated multiple-gene approach has been widely used to reconstruct phylogenetic relationships (37-40). Currently, number of sampled genes seems to be arbitrary. Minimum required genes to resolve phylogentic tree was 20 for 8 yeasts (37) for concatenated multiple gene approach. For the earlier cases, 15 to 50 genes could meet the minimum required genes to get congruent trees with concatenated multiple gene approach and maximum gene-support approach (35, 41). In this study, 43 genes reach the minimum required genes for 6 taxa or less, but not for the 7 taxa in this study. The position of Stramenopiles (a group of eukaryotes) and the relationships among Conosa (amoeba and slime mold), Opisthokonta (fungi and animal) and plant could not be settled by use of more than 100 genes (42). Minimum required genes vary with methods, taxon sets and types of bases.

When a reliable tree is not known, determination of the minimum required genes is difficult. 100% bootstrap support does not mean that the branch is 100% correct. 100% bootstrap support may occur in an alternative branch (43). High bootstrap support does not necessarily signify 'the truth' (44). When a maximum gene-support value is not evidently different, for example, in the case of 7 taxa, it can be recognized that the number of genes used does not meet the requirement of minimum genes. More genes or less taxa is necessary to be re-sampled. This is the outstanding advantage of the maximum gene support tree approach.

In this study, neither consistent results nor evident differences between tree I and tree III are detected in gene support values and taxon jackknife probabilities, which vary with 21 taxon sets, three phylogenetic methods and three gene-scale approaches (Table 2 and 3). However, the clear consensus is that tree II receives the significant lower gene supports than tree I or tree III, and evidently less taxon jackknife probabilities. The significant differences in gene supports are observed by chi-square test (Table 2). We here reject the hypothesis that coelacanth is the closest living relatives of the tetrapods based on these phylogenetic analysis from all 43 genes with all three common phylogenetic methods and all three genome-scale approaches. These results are consistent with previous molecular and palaeontological phylogenetic analysis (3). It was pointed out that the coelacanth is not the closest living relatives of tetrapods based on the latest molecular analysis of two single genes (28). An earlier similar suggestion was proposed based on mitochondrial DNA sequences (16). The recently major published palaeontological studies during the last decade proposed that lungfish (Dipnoi) are the closest living relatives of the tetrapods or alternatively, that coelacanths and lungfish form a monophyletic group that is equally closely related to tetrapods (45- 46). The jackknife probabilities of tree III are slightly higher than that of tree I (Table 2 and 3). It is still very unclear to mention which of lungfish or lungfish-coelacanth sister group is the closest living relative of tetrapods by phylogentic analysis of 43 genes with all three common methods and all three genome-scale approaches. The cause of this puzzle is that the divergence of coelacanth and lungfishes happened in relative short time within a small (20-30 millions years) window in time around 400 million years ago in paleontological data (3, 47) because there was little time and chance for lineage-specific molecular changes to happen, but much time and opportunity for multiple and parallel changes and their accumulation since the origin of these two lineages (3). So, it is most difficult to discriminate against them using ad-hoc molecular phylogenetic methods and algorithms when available sequence data of genes are currently limited. We can not make decision of accepting hypotheses 1 or 3 based on current evidences of this study. More genes still need to be involved in order to resolve this question in the future. Additionally, new phylogenetic methods and geneome-scale algorithms may be helpful and need to be developed. Some researchers consider NJ as phenetic technique, we include it here to demonstrate that neither phenetic nor phylogenetic analyses reliably recover a coelacanth/tetrapod pairing.

In conclusion, we rejected hypothesis 2 that the coelacanth is the closest living relatives of tetropads due to its significant low gene supports and low jackknife probabilities based on the phylogenetic analysis results using 43 genes with all three common phylogenetic/phenetic methods and three genome-scale approaches. However, determination of hypothesis 1 and 3 requires to be further studied in the future.

## Materials and Methods

**Sequence Collection.** The sequences of encoding amino acid residues of 43 genes were mined from GenBank using the datamining tool called as BLAST. Having been compared from the supplementary materials (33), these sequences of 43 genes were previously analyzed using the genome-scale approach of concatenated genes although the lengths of some sequences were different. One gene (FSCN1) was not included because some taxa lacked the FSCN1 sequence in GenBank (33). In order to compare the results with the genome-scale concatenated gene approach (33), the same 7 taxa were included: Mammal (M), Bird (B), Amphibian (A), Coelacanth (C), Lungfish (L), Ray-finned Fish (R), and Shark (S). Amino acid sequences were used for phylogenetic analysis.

**Phylogenetic Analysis.** Sequences of an individual gene were aligned using ClustalX with default settings (48). All alignments of single genes were manually edited to exclude insertions or deletions and uncertain positions from further analysis.

The phylogenetic analysis software PAUP* (Version 4.0b10) (49) was used for tree inference with MP method. Each set of sequences of single genes or concatenated genes was analyzed under the optimality criteria of maximum parsimony for MP. The MP analyses were performed with unweighted parsimony. The sequences also were analyzed with ML and NJ using default settings by PHYLIP (36).

**Approach of Concatenated Genes.** The first step is to concatenate small alignments of single genes into one large alignment, and then a tree is reconstructed using the large alignment (43, 50-53). The bootstrap consensus tree was searched using the branch-and-bound algorithm for MP, and the full heuristic search was used for NJ and ML based on a 50% majority rule. One thousand replicates were used for all tests except for ML, where 100 replicates were completed.

**The Maximum Gene-support Tree Approach.** All single gene trees were recovered using all 43 individual genes using MP, ML and NJ methods. Tree distances for all pairwise comparisons among trees were calculated using the symmetric difference metric by PAUP* (36, 49). This distance is the number of steps required to convert between two trees, that is, the number of branches that differ between a pair of trees (54). Two trees with identical topology have a tree distance of zero. A maximum gene-support tree was defined as a unique tree that was recovered by the most genes of all these used genes (35). A computer program in C language for calculating gene-support is also available from the authors upon request (shan@cs.dal.ca).

**The Maximum Gene-support Branch Approach.** Based on all single gene trees recovered using all 43 individual genes, majority-rule consensus tree with less than 50% parameter setting was calculated by PAUP* (49). Gene support was obtained for each branch by corresponding support values.

**Taxon Jackknife Sub-sampling.** We used a jackknife approach to sub-sample 6, 5, 4 taxa from 7 taxa each time to reduce alternative trees because the number of alternative trees increases exponentially with the number of taxa (6→945, 7→10,395, n→ (2n-3)!!) (36). The debate of taxon sampling has not terminated. On the hands, the accuracy was enhanced dramatically with the addition of taxa (54). On one other hands, adding taxa can reduce accuracy and increase the probability of distorting the tree topology (56). Adding characters can always increase the accuracy (54-56). So, genes should be included as many as possible. Sequence data of 43 genes that are all currently available in GenBank were used in this study.

**Chi-square Test.** Statistical significant difference between the gene supports of tree II and tree I / III was analyzed by means of chi-square test.

**ACKNOWLEDGMENTS.** We thank Richard Winterbottom for critically reading an earlier draft, helpful comments and valuable suggestions, and Adam Aspinall for reading and valuable comments. This work is partially supported by the NSERC grant ORGPIN 341854, the CRC grant 950-2-3617 and the CFI grant 203617.**References**

Figures:

a. Tree I: Lungfish hypothesis
b. Tree II: Coelacanth hypothesis
c. Tree III: Lungfish-coelacanth sister grouping hypothesis
d. Tree IV: Lungfish-coelacanth-tetrapod trichotomy

**Fig. 1.** Four alternative phylogenetic trees among tetrapod, coelacanth and lungfish lineages.

**Fig. 2.** Four alternative types of phylogenetic trees of 7 taxa.

Tables:

Table 1 Tree types of seven taxa with three methods and three genome-scale approaches

Table 2 Gene supports of four tree types, and six, five and four taxon sets inferred with MP, ML and NJ

Table 3 Bootstrap supports, gene supports and taxon jackknife probabilities of four tree types, and six, five and four taxon sets using three approaches recovered with MP, ML and NJ

# Fig. 1

## a. Tree I

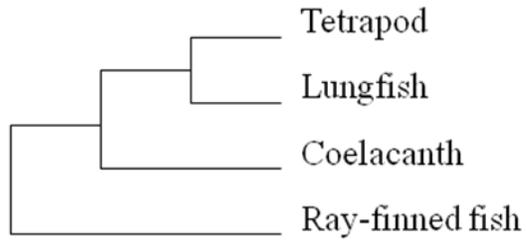

## b. Tree II

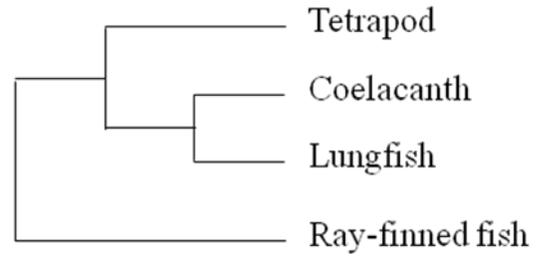

## c. Tree III

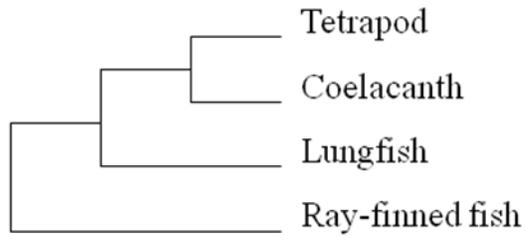

## d. Tree IV

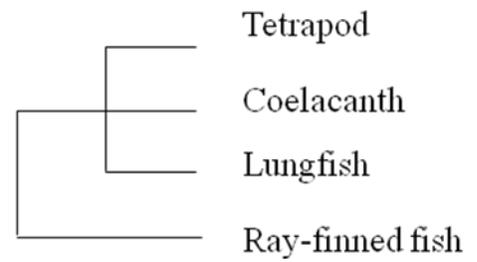

1
2

a.

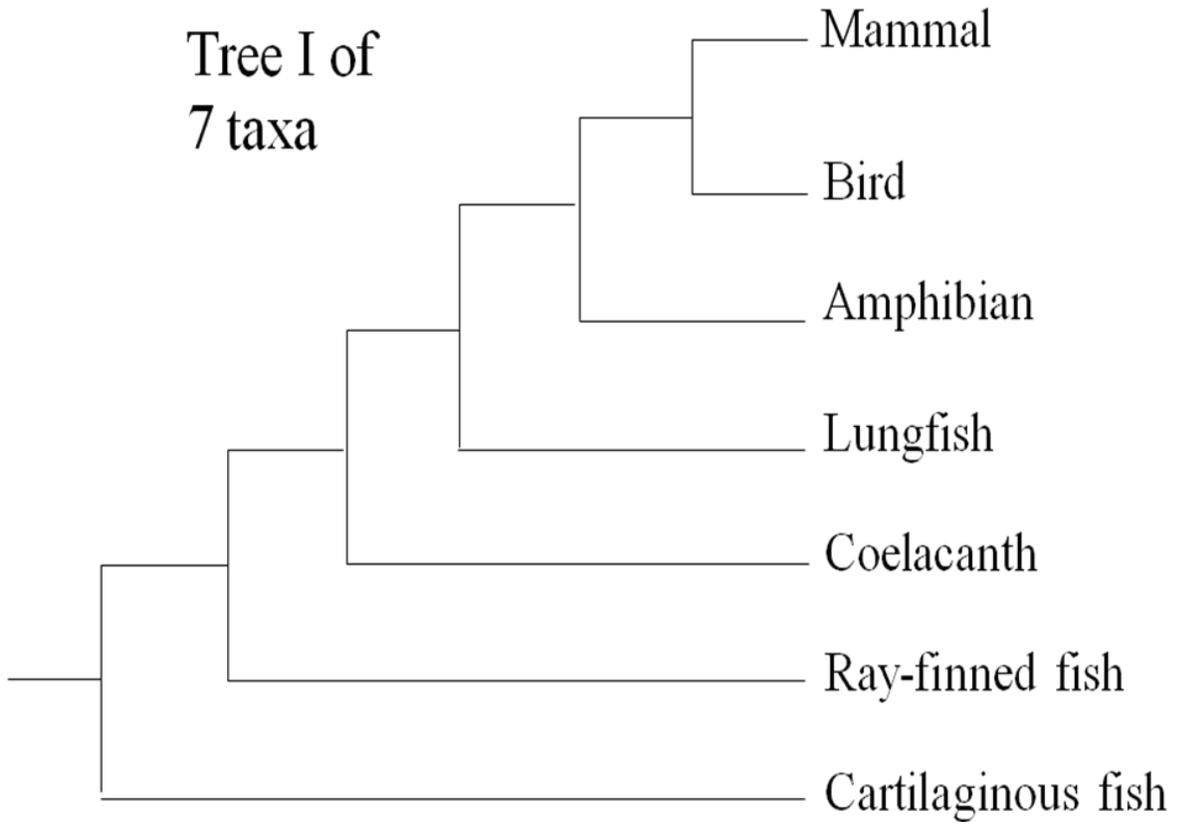

b.

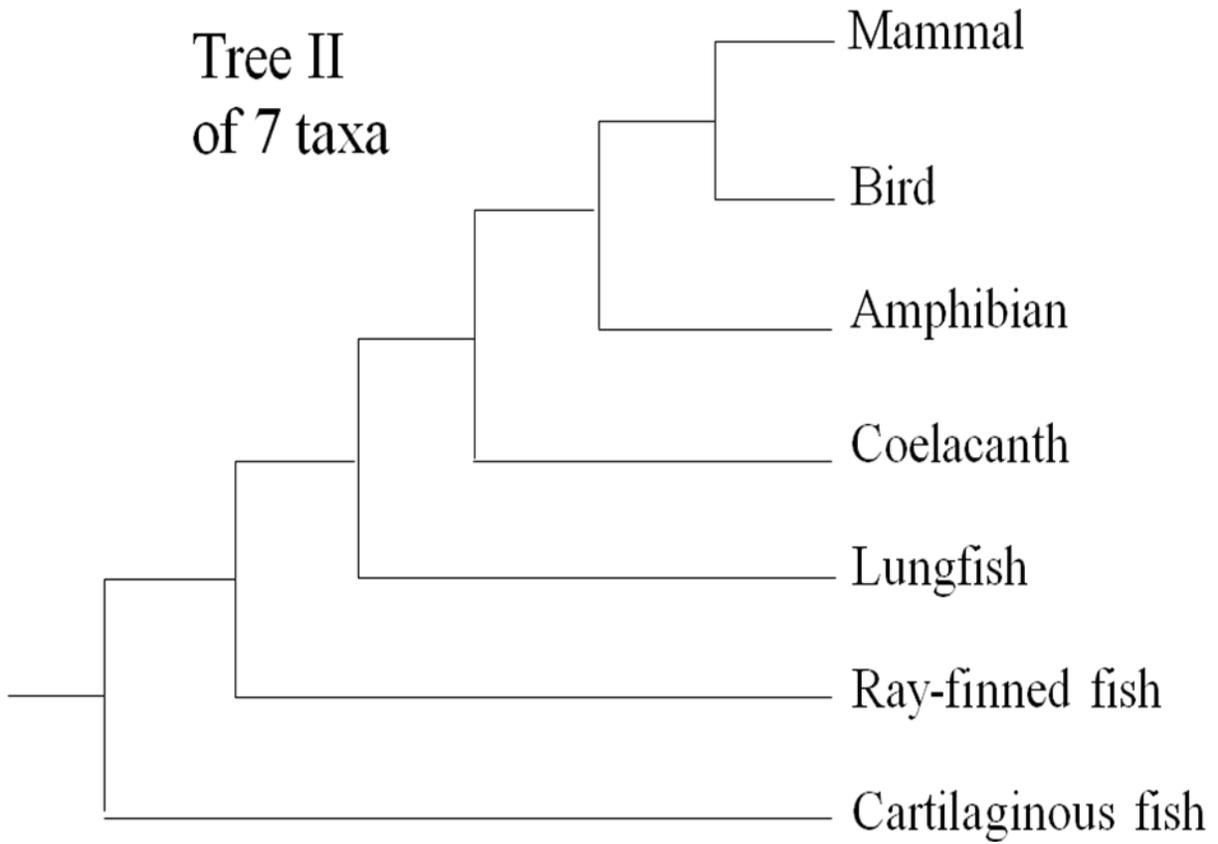

c.

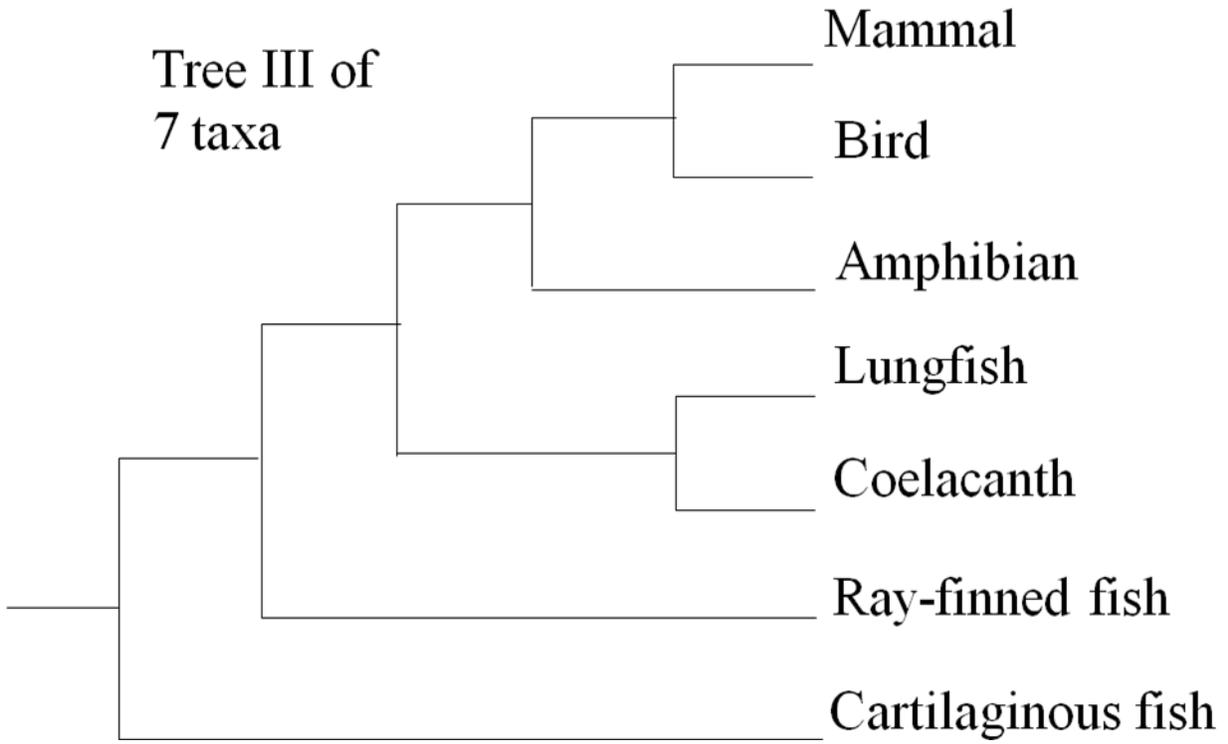

d.

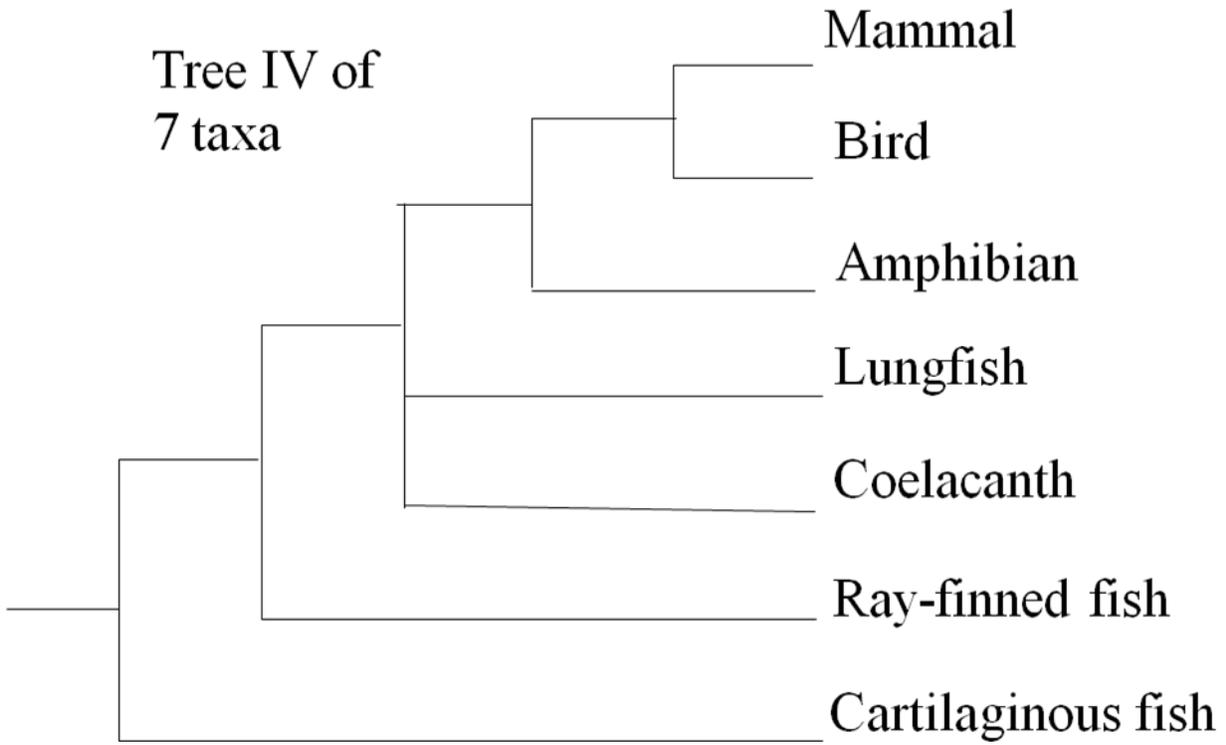

**Fig. 2.**

Tables:

Table 1 Tree types of seven taxa with three methods and three genome-scale approaches

| Genome-Scale Approaches | Phylogenetic Methods | | |
|---|---|---|---|
| | MP | ML | NJ |
| Concatenation | I(67%) | I(100%) | III(87%) |
| Maximum Gene Support Tree | I(2)/II(2)/III(2)/IV(2) | I(2)/II(1)/III(2)/IV(0) | I(2)/II(1)/III(1)/IV(0) |
| Maximum Gene Support Branch | III(9) | III(9) | II(10) |

Notes: The numbers in ( ) are gene supports for maximum gene support tree and maximum gene support branch approaches, bootstrap supports (%) for concatenation approach, respectively.

Table 2 Gene supports of four tree types, and six, five and four taxon sets inferred with MP, ML and NJ

| Taxon Set | Type of Trees | | | | | | | | | | | |
|---|---|---|---|---|---|---|---|---|---|---|---|---|
| | MP | | | | ML | | | | NJ | | | |
| | Tree I | Tree II | Tree III | Tree IV | Tree I | Tree II | Tree III | Tree IV | Tree I | Tree II | Tree III | Tree IV |
| Six taxon sets: | | | | | | | | | | | | |
| BACLRS | 2 | 3 | 3 | 0 | 3 | 1 | 5 | 0 | 4 | 3 | 3 | 0 |
| MACLRS | 1 | 1[+] | 6 | 0 | 3 | 3 | 2 | 0 | 1 | 1 | 2 | 0 |
| MBACLR | 6 | 1[+] | 6 | 0 | 4 | 3 | 5 | 0 | 3 | 3 | 8 | 0 |
| MBACLS | 4 | 3 | 5 | 0 | 6 | 3 | 4 | 0 | 2 | 7 | 5 | 0 |
| MBCLRS | 5 | 3 | 7 | 1 | 5 | 4 | 5 | 0 | 7 | 3 | 3 | 0 |
| Five taxon sets: | | | | | | | | | | | | |
| ACLRS | 6 | 5 | 6 | 0 | 8 | 4 | 3 | 0 | 7 | 4 | 3 | 0 |
| BACLR | 6 | 6 | 8 | 0 | 7 | 4 | 10 | 0 | 8 | 4 | 10 | 0 |
| BACLS | 9 | 4 | 8 | 0 | 13 | 4[*] | 6 | 0 | 10 | 9 | 6 | 0 |
| BCLRS | 8 | 4 | 6 | 0 | 8 | 5 | 5 | 0 | 8 | 7 | 3 | 0 |
| MACLR | 7 | 5 | 9 | 0 | 6 | 6 | 9 | 0 | 5 | 4 | 10 | 0 |
| MACLS | 4 | 5 | 10 | 0 | 9 | 3[+] | 8 | 0 | 3 | 9 | 8 | 0 |
| MBCLR | 10 | 4[*] | 14 | 0 | 12 | 8 | 15 | 0 | 12 | 8 | 15 | 0 |
| MBCLS | 10 | 11 | 8 | 0 | 14 | 6[+] | 12 | 0 | 15 | 10 | 9 | 0 |
| MCLRS | 5 | 6 | 9 | 0 | 5 | 4 | 6 | 0 | 6 | 4 | 3 | 0 |
| Four taxon sets: | | | | | | | | | | | | |
| ACLR | 13 | 13 | 16 | 1 | 16 | 12 | 15 | 0 | 9 | 10[*] | 24 | 0 |
| ACLS | 14 | 14 | 13 | 2 | 16 | 13 | 14 | 0 | 14 | 12 | 17 | 0 |
| BCLR | 19 | 11 | 13 | 0 | 15 | 11 | 17 | 0 | 16 | 16 | 11 | 0 |
| BCLS | 18 | 11 | 13 | 1 | 21 | 8[*] | 14 | 0 | 22 | 8[*] | 13 | 0 |
| MCLR | 12 | 13 | 17 | 1 | 12 | 15 | 16 | 0 | 10 | 14 | 19 | 0 |
| MCLS | 11 | 14 | 16 | 2 | 13 | 13 | 17 | 0 | 13 | 13 | 17 | 0 |

Notes: The taxa included: Mammal (M), Bird (B), Amphibian (A), Coelacanth (C), Lungfish (L), Ray-finned Fish (R), and Shark (S). [+, *] indicated chi-square test significant level at $P < 0.10, 0.05$ between the gene-supports of tree II and tree I/III, respectively.

Table 3 Bootstrap supports, gene supports and taxon jackknife probabilities of four tree types, and six, five and four taxon sets using three approaches recovered with MP, ML and NJ

| | Approaches and Methods | | | | | | | | |
|---|---|---|---|---|---|---|---|---|---|
| | MP | | | ML | | | NJ | | |
| Taxon Set | CT | MGT | MGB | CT | MGT | MGB | CT | MGT | MGB |
| Six taxon sets: | | | | | | | | | |
| BACLRS | I (73%) | III (3) /II (3) | III (9) | I (53%) | III (5) | III (11) | III (79%) | I (4) | I (9) |
| MACLRS | AT (n/a) | III (6) | III (11) | I (47%) | III (3)/II (3) | I (17) | III (95%) | III (2) | III (6) |
| MBACLR | III (86%) | III (6) /I (6) | III (12) | I (100%) | III (5) | III (16) | III (96%) | II (7) | III (21) |
| MBACLS | I (80%) | III (5) | III (11) | I (53%) | I (6) | I (20) | III (69%) | III (8) | III (9) |
| MBCLRS | I (61%) | III (7) | III (14) | I (50%) | III (5)/I (5) | III (9) | III (85%) | I (7) | I (7) |
| Five taxon sets: | | | | | | | | | |
| ACLRS | III (52%) | III (6)/I (6) | I (11) | I (32%) | I (8) | III (11) | III (92%) | I (7) | I (9) |
| BACLR | III (85%) | III (8) | II (20) | III (48%) | III (10) | III (15) | III (95%) | III (10) | II (12) |
| BACLS | I (100%) | I (9) | I (21) | I (22%) | I (13) | I (21) | I (100%) | I (10) | I (12) |
| BCLRS | I (87%) | I (8) | I (18) | I (46%) | I (8) | I (14) | III (59%) | I (8) | I (9) |
| MACLR | III (94%) | III (9) | III (21) | III (52%) | III (9) | III (16) | III (99%) | III (10) | III (12) |
| MACLS | I (100%) | III (10) | III (19) | I (46%) | I (9) | III (12) | III (81%) | II (9) | III (12) |
| MBCLR | I (73%) | III (14) | III (32) | II (43%) | III (15) | III (16) | III (88%) | III (15) | III (13) |
| MBCLS | I (100%) | II (11) | II (29) | II (40%) | I (14) | I (32) | III (75%) | I (15) | II (13) |
| MCLRS | III (53%) | III (9) | III (13) | III (45%) | III (6) | II (15) | III (96%) | I (6) | III (9) |
| Four taxon sets: | | | | | | | | | |
| ACLR | III (95%) | III (16) | III (17) | III (55%) | I (16) | III (16) | III (99%) | III (24) | III (20) |
| ACLS | AT (n/a) | I/II (14) | I (14) | I (100%) | I (16) | II (17) | IV (49%) | III (17) | I (13) |
| BCLR | III (51%) | I (19) | I (19) | I (51%) | III (17) | III (17) | III (75%) | I (16) or II (16) | I (19) |
| BCLS | III (80%) | I (18) | I (18) | I (46%) | I (21) | I (21) | I (64%) | I (22) | I (15) |
| MCLR | III (80%) | III (17) | III (17) | III (49%) | III (16) | III (16) | III (93%) | III (19) | III (17) |
| MCLS | III (61%) | III (16) | III (16) | III (50%) | III (17) | III (18) | III (92%) | III (17) | III (16) |
| JKF: | I (40%) | I (27.5%) | I (30%) | I (60%) | I (47.5%) | I (30%) | I (10%) | I (42.5%) | I (40%) |
| | II (0) | II (10.0%) | II (10%) | II (10%) | II (2.5%) | II (10%) | II (0) | II (12.5%) | II (10%) |
| | III (50%) | III (62.5%) | III (60%) | III (30%) | III (50%) | III (60%) | III (85%) | III (45%) | III (50%) |

| | | | | | | | | |
|---|---|---|---|---|---|---|---|---|
| IV (0) | IV (0) | IV (0) | IV (0) | IV (0) | IV (0) | IV (5%) | IV (0) | IV (0) |
| AT (10%) | AT (0%) | AT (0) | AT (0) | AT (0) | AT (0) | AT (0) | AT (0) | AT (0) |

___________________________________________________________________________

**Notes:** CT = Concatenation tree; MGT = Maximum gene-support tree; MGB = Maximum gene-support branch; JKF = Taxon jackknife probabilities (%), the number in ( ) is gene support for MGT and MGB, bootstrap support (%) for CT; AT = alternative tree; n/a = not available